\documentclass[12pt, preprint]{aastex}

\slugcomment{Submitted to Astrophys. J. Let.}

\shorttitle{Evolution of Triton}
\shortauthors{\' Cuk and Gladman}

\begin{document}

\title{Constraints on the Orbital Evolution of Triton} 

\author{Matija \' Cuk\altaffilmark{1} and Brett J. Gladman\altaffilmark{1}}
\altaffiltext{1}{Department of Physics and Astronomy, University of British Columbia,
    6224 Agricultural Rd, Vancouver, B.C. V6T 1Z1, Canada, E-mail:cuk@astro.ubc.ca}

\begin{abstract}

We present simulations of Triton's post-capture orbit that confirm the importance of Kozai-type oscillations in its orbital elements. In the context of the tidal orbital evolution model, these variations require average pericenter distances much higher than previously published, and the timescale for the tidal orbital evolution of Triton becomes longer than the age of the Solar System. Recently-discovered irregular satellites present a new constraint on Triton's orbital history. Our numerical integrations of test particles indicate a timescale for Triton's orbital evolution to be less than $10^5$ yrs for a reasonable number of distant satellites to survive Triton's passage. This timescale is inconsistent with the exclusively tidal evolution (time scale of $>10^8$ yrs), but consistent with the interestion with the debris from satellite-satellite collisions. Any major regular satellites will quickly collide among themselves after being perturbed by Triton, and the resulting debris disk would eventually be swept up by Triton; given that the total mass of the Uranian satellite system is 40\% of that of Triton, large scale evolution is possible. This scenario could have followed either collisional or the recently-discussed three-body-interaction-based capture.
\end{abstract}


\section{Introduction}


Neptune's Triton is the only large planetary satellite to orbit retrograde relative to the planet's rotation. \citet{mcc66} and later \citet{mck84} suggested that Triton is a captured satellite, whose originally eccentric orbit was circularized due to tidal dissipation within Triton. \citet{gol89} postulate that Triton was captured from heliocentric orbit by a collision with a pre-existing satellite, and its initial high-eccentricity orbit then evolved due to tidal dissipation alone. They showed that the tidal evolution timescale is significantly shorter than the age of the Solar System (a few times $10^8$ years), even when the variations in Triton's eccentricity with $1/2$-yr period are accounted for. However, \citet{mck95} noted that the criteria for non-disruptive capture are much stricter than \citet{gol89} calculated. If the original regular satellites of Neptune and Uranus were similar, a collision with the largest moons (preferred due to their cross-sections) would disrupt Triton, with re-accretion on an orbit inclined to Neptune's equator being impossible. 


\citet{ml95} suggested instead that Triton was captured and its orbit was evolved by aerodynamic drag in Neptune's primordial protosatellite nebula, and that after its orbit ewas circularized Triton's gravity would be sufficient to clear a disk gap and thus halt further evolution. Gas drag has been suspected as a capture mechanism for small and distant irregular satellites, but capture of Triton would require unprecedented gas densities, requiring very close approaches to Neptune.


\citet{ah04} propose a three-body capture scenario for Triton. They suggest pre-capture Triton may have been a member of a binary, whose disruption during a Neptune encounter led to Triton's capture and its companion's escape. Their work addresses only the capture itself, leaving the physics of post-capture evolution unchanged. 

\section{Effect of Solar Perturbations}


Any capture mechanism, be it collision, gas drag or 3-body interaction, is likely to introduce Triton on a large, highly eccentric orbit.  Distant satellite orbits are perturbed primarily by the Sun, inducing precession of angular variables and oscillations in eccentricity and inclination, with minor semimajor axis variations. The two important periodic perturbations are those associated with $1/2$ of the planet's orbital period (``evection'') and $1/2$ of the precession period of the argument of pericenter $\omega$ (``Kozai behavior''). For early Triton, evection was first noted by \citet{gol89}; conservation of orbital angular momentum during tidal evoution implies initial pericenter of 7 Neptune radii ($R_N$; Triton's present orbit has $a=14 R_N$), and evection-induced oscillations in $e$ produced minimum pericenters of $5 R_N$. 

\begin{figure*}
\includegraphics[angle=270, scale=.35]{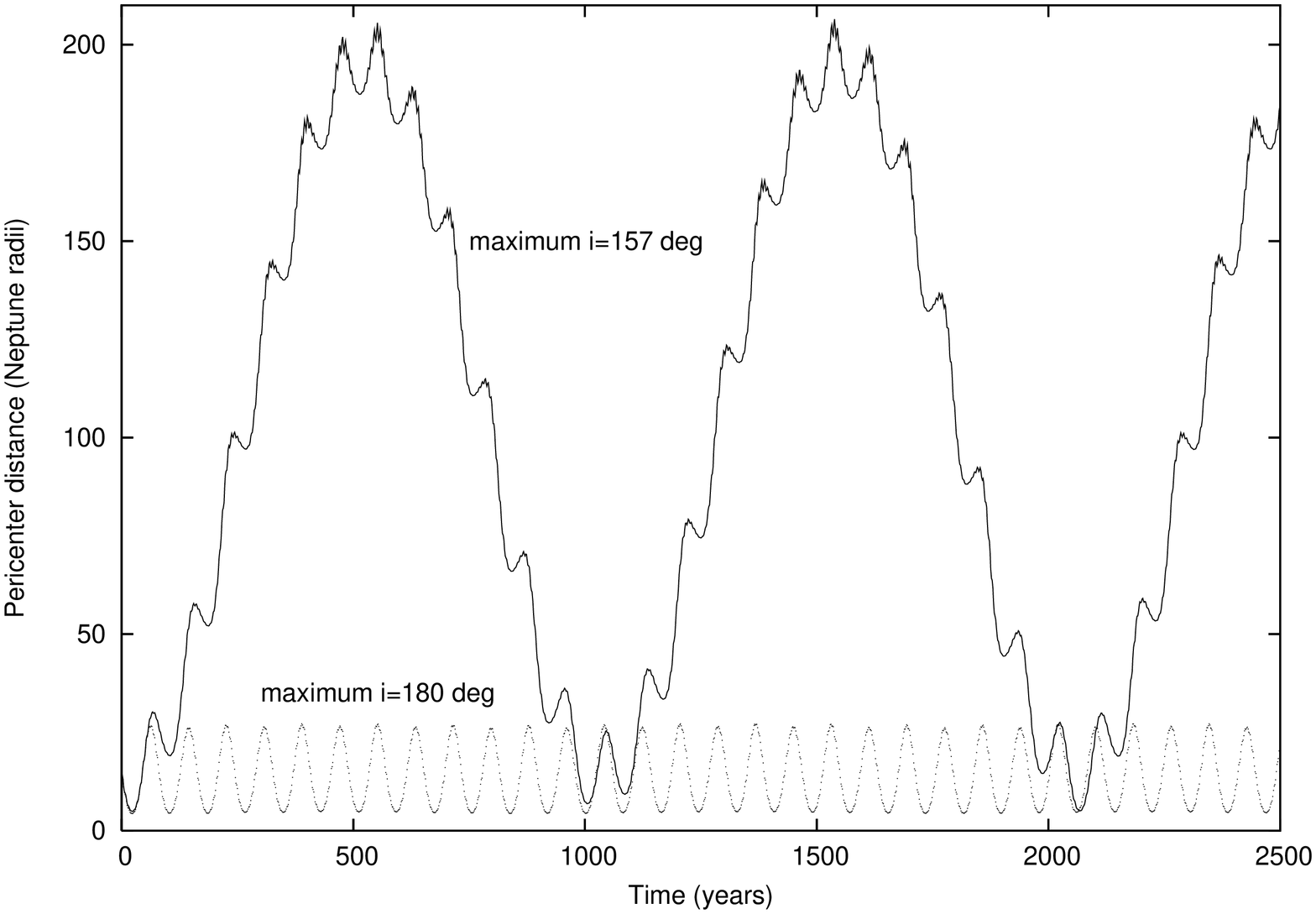}  
\caption{Pericenter distances of Triton on large and eccentric post-capture orbits, with $i_{min}=157^{\circ}$ (at $\omega=90^{\circ}$; solid line) and $i=180^{\circ}$ (dotted line). When the pericenter is furthest from neptune, the inclination (with respect to the Neptune's orbit) is at the minimum value ($i_{min}=110^{\circ}$ for $i_{max}=157^{\circ}$).}
\label{peric}
\end{figure*}

However, \citet{gol89} ignored Kozai oscillations in $e$, which must have been present if Triton's inclination $i$ at closest approach was same as now ($i=157^\circ$, measured with respect to Neptune's equator). Kozai osillations require that when $\omega=90^{\circ}$ or $270^{\circ}$, both $e$ and $i=157^{\circ}$ are at a maximum ($e$ and $i$ oscillate in phase for retrograde satellites). Since almost all tidal dissipation occurs during this high-$e$ phase of the Kozai cycle (when the pericenter distance is smallest), this $157^\circ$ inclination will be conserved as the maximum one for the Kozai cycle, while the minimum one (coinciding with $\omega=0^{\circ}$ and $180^{\circ}$) will be affected by dissipation. Using more complete treatment of tides, \citet{chy89} show that tides raised on Triton could not have led to a significant inclination change as long as it was in Cassini state 1. Trapping in the much less likely Cassini state 2 would have caused a rapid increase in its inclination (i.e., closer to $180^{\circ}$), rather than a decrease \citep{jan89}, so Triton's inclination relative to the local Laplace plane was always $<157^{\circ}$. This assumes orbital evolution slower than the nodal precession, preserving inclination relative to the local Laplace plane, which would initially be close to Neptune's orbital plane but would subsequently move closer to Neptune's equator. This condition is true for the tidal model \citep{gol89} but not the gas-drag model \citep{ml95}. Fig. \ref{peric} presents two illustrative short orbital evolutions. Using initial conditions of $a=2 \times 10^7$~km$=800 R_N=0.18 R_H, e=0.982, \omega=90^{\circ}$, and varying inclinations, we re-create two possible post-capture orbits for Triton. Both evection ($P \simeq 80$ yrs) and Kozai oscillations ($P \simeq 1000$ yrs) are clearly visible in the evolution of the inclined orbit, while the $i \approx 180^{\circ}$ case, shows only evection-related oscillations (whose amplitudes are in agreement with results from \citet{gol89} Fig. 2 for $a=800 R_N$). However, if tidal dissipation alone evolved Triton's orbit, only the inclined case can lead to present inclination of $157^{\circ}$. 

This conclusion points to a paradox. \citet{gol89} modelled the tidal orbital evolution with the standard relation:
\begin{equation}
\label{goldreich}
{1 \over a}{da \over dt}=-{21 \over 64} {n \over \mu}{a R_T^5 \over q^6}{k_2 \over Q},
\end{equation}
where $a, t, q, n, k_2$ and $Q$ are respectively the semimajor axis, time, pericenter distance, mean motion, tidal Love number and tidal disspation factor, and $R_T=1353$~km is Triton's radius. We numerically averaged Eq. \ref{goldreich} over several full Kozai periods, based on the output of the integration shown in Fig. \ref{peric}. Using $k_2=0.1$ and $Q=100$ for Triton \citep{gol89}, the resulting timescale for $1/e$ reduction in $a$ is 3.5 Gyr for the inclined orbit and 0.11 Gyr for the near-coplanar orbit (cf. Goldreich et al. Fig.1). The requirement of an inclined orbit means that tides alone are not capable of circularizing Triton's orbit.

\section{Interaction with Possible Regular Satellites}

Neptune's satellite system predating Triton's capture was likely similar to that of Uranus, as the planets are virtual twins in many important characteristics. Uranus possesses five sizeable satellites at $5R_U < a < 23R_U$,  with total mass $\approx$40\% of Triton's. It is likely that Triton's capture led to destruction or ejection of any existing regular satellites outside $5 R_N$. 

To explore the fate of these moons, we integrated five massless particles under the influence of Triton on the $i=157^{\circ}$ orbit of Fig. \ref{peric}. The innermost satellite was put at $a_1=4.8 R_N$, and the others on $a_i=1.5 a_{i-1}$; all the orbits were circular and in the plane of Neptune's equator (using current values of Neptune's obliquity and $J_2$). We found that the orbital crossing of the pairs 4-5 and 3-4 only a few centuries into the simulation. To calculate the collisonal timescale, we assigned our three outermost particles the radii of Umbriel, Titania and Oberon, and we took the satellites to be uniformly distibuted over all radii $a(1-e)<r<a(1+e)$, and latitudes $-i < \beta <+i$. Under these (admittedly rough) approximations, we estimated the collision probability for every output interval in our simulation and kept track of the cumulative collision probability for each pair. In both cases, the cumulative probability reaches unity at about 1000 years. While our approximations likely underestimate collision timescales, a more sophisticated method would not raise the result more than a factor of a few. Due to its long orbital period (about 7 yrs), most of it spent far from Neptune, eccentric Triton would be unlikely to collide with any of the satellites on such short timescales.

The collision would likely destroy both moons and create a debris ring around Neptune. This debris would rapidly disrupt the remaining regular satellites, grinding them down to small pieces incapable of destroying Triton. The resulting massive disk out to about $20 R_N$ would interact strongly with Triton, whose perturbations would prevent re-accretion of the disk particles from taking place (cf. Banfield \& Murray, 1992). The mutual interaction of the retrograde Triton and prograde disk, either through tidal torques or direct collisions, would cause decay of Triton's angular momentum; here we focus on the latter. 

To find the approximate evolution timescale of Triton's orbit due to the debris disk, we modeled the disk as having uniform surface density and extending out to $20 R_N$, and then integrated Triton's orbit giving it a instantaneous ``kick'' every time it goes through Neptune's equatorial plane within $20 R_N$. Like \citet{ml95}, we find that Triton's lifetime depends largely on its inclination: inclined orbits survive much longer than coplanar ones. Our timescales for the decay of orbits with $i=157^{\circ}$ at $\omega=90^{\circ}$ are on the order of a few times $10^5$~yrs, while the orbit with $i=175^{\circ}$ at $\omega=90^{\circ}$ would have an $e$-folding time of a few times $10^4$~yrs. While our disk model and resolution of the passages are crude, the robust conclusion is that the debris-drag timescale for orbital evolution is much shorter than the tidal one. Although our disk is much less massive than that of \citet{ml95} (who use a minimum mass nebula) Triton can still evolve quickly due to deeper penetration into the disk. Finally, since the disk is less massive than Triton, there is a natural end to this stage of the evolution, so there is no need for specific mechanisms ensuring Triton's survival. 

If Triton's evolution had been this fast ($10^4-10^5$~yrs), then the assumption of conservation of inclination relative to local Laplace plane is incorrect, as the nodal precession timescale (at the boundary between an inner region dominated by Neptune's oblatneness and an outer region dominated by the Sun) is longer than $10^5$~yrs (the transition distance depends on $J_2$ as the fifth root, making the influence of the additional material from the debris disk modest). Therefore, the pre-transition inclination of Triton could have been anything from $128^{\circ}$ to $174^{\circ}$ if the evolution was fast. Because passages through the disk would at that point ($a \simeq 80-100 R_N$) happen for a wide range of values for $\omega$ (and not only $\omega \simeq 90^{\circ}$ when $e$ is high), the constraints on the initial inclination are very weak. 

\section{Interaction with Irregular Satellites}

Nereid and the other five now-known irregular satellites of Neptune \citep{hol04} can help us constrain Triton's post-capture orbit. During its early orbital history, Triton's orbit would intersect those of the irregulars, making gravitational scattering possible. There is reason to think that Neptune's irregulars had their orbits modified after capture: almost all irregular satellites of the other three giant planets have pericenters $q$ around 100 and 200 planetary radii, in case of direct and retrograde satellites, respectively. Four outer neptunians do not approach the planet closer than about $350 R_N$, while S/2002~N1 and Nereid come within about $50 R_N$; the latter two objects will be addressed in the next section. 

\begin{figure*}
\includegraphics[angle=270, scale=.35]{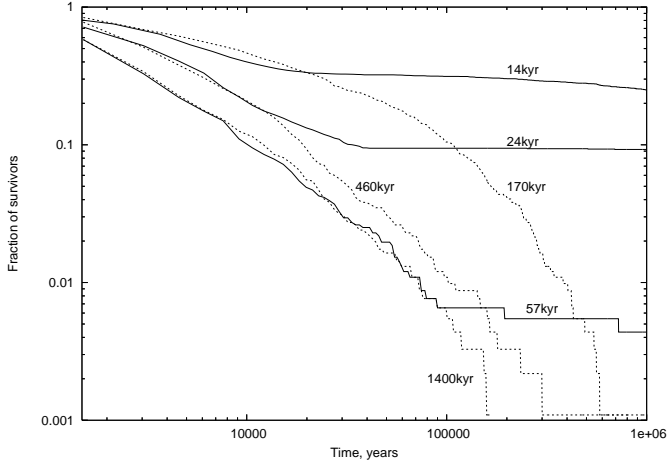}  
\caption{Evolution of the fraction of surviving test particles over time. The labels on the graph give time Triton takes to evolve to half its initial $a$ in each simulation. Dashed lines plot evolution with Triton's initial $i_0=157^{\circ}$ and $a=10, 20, 30\times 10^6$~km (right to left), while the solid lines plot simulations with $i_0=175^{\circ}$ with $a=5, 10, 20 \times 10^6$~km (top to bottom). Neptune's hill sphere is $110 \times 10^6$~km in radius.} 
\label{lt}
\end{figure*}

Can large-$q$ irregulars result from Triton's passage?  Fig. \ref{lt} shows six simulations using a symplectic integrator that includes Triton evolving under the influence of the disk while interacting with a swarm of test particles representing ``original'' irregulars. The 2044 test particles were originally put at $a=100-1820R_N$, $i=0-180^{\circ}$,  $\omega=90^{\circ}$ and with fixed $q=100~R_N$ (for direct) or $q=200~R_N$ (retrograde orbits). Triton's initial conditions were varied (see Fig. \ref{lt} caption). The curves in Fig. \ref{lt} give the fraction of surviving bodies over the 1~Myr integrations, normalized to the number of survivors when Triton is not present (some particles are inherently unstable).

Fig. \ref{lt} shows that a significant number of irregulars survives Triton's passage only if the duration of the interaction is below $\sim 10^5$~yrs. In such cases, Triton's apocenter drops below the inner edge of the irregular population, and the depletion stops. In these models such rapid evolution occurs only for $i_{max}=175^{\circ}$, for then Triton's pericenter is almost continuously within the disk. An orbit with this maximum inclination will occur for a large fraction of $e>0.9$ post-capture orbits. Also, requirement for $\tau < 10^5$~yrs practically excludes tidal evolution {\it independently} from arguments in section 2.



\begin{figure*}
\includegraphics[angle=270, scale=.35]{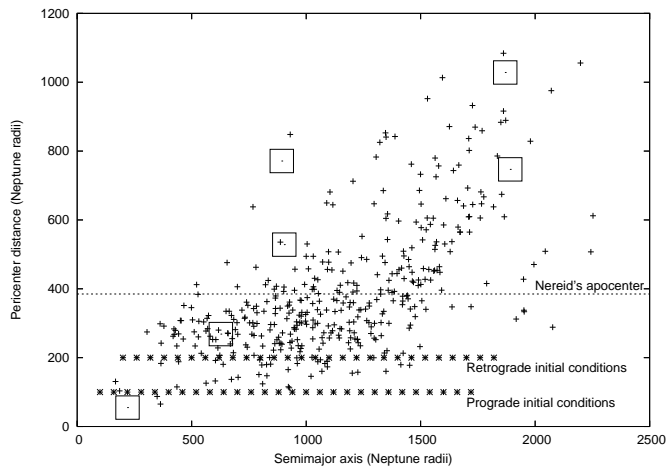}  
\caption{Semimajor axis--pericenter plot for simulated test particles (initial conditions:asterisks, final orbits:pluses) and real satellites (large boxes). Nereid's apocenter distance is shown as a dashed line.}
\label{aqplot}
\end{figure*}

Fig. \ref{aqplot} compares the semimajor axes and pericenters of Nereid and the five other irregulars with the surviving test particles in the simulation represented by the topmost curve in Fig. \ref{lt}. The scattered particles bridge the gap between initial conditions corresponding to other planetary irregulars and Neptune's. However, the mechanism leaves many satellites at $q=200-400 R_N$. Interestingly, Nereid's apocenter distance seems to separate the bulk of the surviving particles and the observed moons. To test the importance of scattering by Nereid on the irregulars, we ran two simulations similar to the ones already described, except that now non-evolving Nereid was the scatterer and the simulations lasted $10^6$ and $10^7$ years, respectively. There was a noticeable effect on the stability of particles, and some particles in the center of the ``stable zone'' were observed to escape. Nereid's escape speed ($v_{esc}=0.25 v_{orb}$ at apocenter) apparently makes it an efficient scatterer, perhaps allowing the moons on orbits permanently out of Nereid's reach to dominate the surviving population. 


\section{Early Dynamics of Nereid-Type Objects}

Nereid orbits between 55 and 385$R_N$, and has been previously suggested to be a regular satellite scattered to its current orbit by Triton. We report preliminary findings on the secular orbital evolution of Nereid-like objects just after the end of close encounters with Triton. By ``Nereid-like'' we mean that the precession is dominated by Triton, rather than the Sun. The radius of the zone within which the planetary oblateness dominates solar perturbations is $r_c=[2 J_2 (M_N/M_{Sun}) R_N^2 a_N^3]^{1/5}$, where $M_N$ and $M_{Sun}$ are the masses of Neptune and the Sun, and $a_N$ is the planet's semimajor axis. We estimate the importance of Triton by putting $J_2=(M_T R_N^2)/(2 M_N a_T^2)$, where subscript $T$ refers to Triton. If we take $a_T \simeq 1/2 r_c$, we find $r_c=(0.25 M_T/M_{Sun})^{1/3} a_N \simeq 8 \times 10^6$~km. Therefore, ``Nereid-type'' objects are those orbiting within about 7\% of Neptune's Hill sphere (for Nereid, $a=5.5 \times 10^6$~km$=220 R_N$). 

\begin{figure*}
\includegraphics[angle=270, scale=.35]{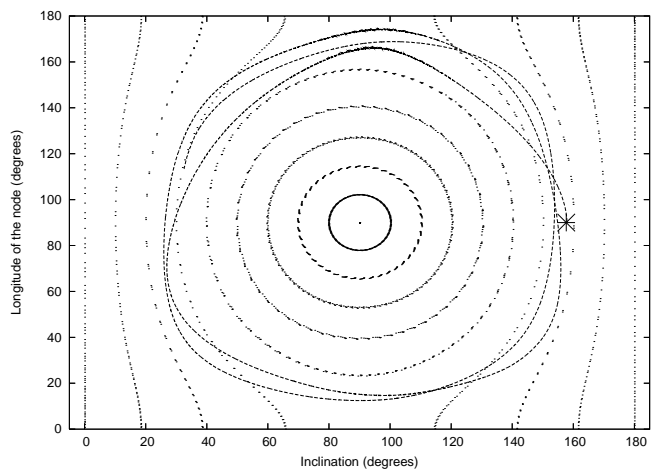}  
\caption{Evolution of early-Nereid-type orbits in $i$-$\Omega$ space: dotted curves represent results of an analytic model while the solid curve shows the history of a numerically integrated particle ($a_2=1.78a_T$; $e_T=0.6$ in both cases). The asterisk marks the initial condition. The inclination is measured with respect to Triton's orbital plane. The phase space for $\Omega=180^{\circ}-360^{\circ}$ is identical, with another libration island at $\Omega=270^{\circ}, i=90^{\circ}$.}
\label{ioplot}
\end{figure*}

The most important secular interaction between two non-intersecting orbits is through the quadrupole term in the interaction Hamiltonian (i.e. one containing $a_1^2/a_2^3$; subscripts 1 and 2 refer to inner and outer body, respectively). The disturbing function for the outer body is \citep{inn97}:
\begin{equation}
\label{innanen}
R={G M_T a_T^2 \over 8 b_2^3} [ 2 + 3 e_T^2 - (3 +12 e_T^2 -15 e_T^2 \cos^2 \omega_T) \sin^2 i_2 ].
\end{equation}
where $b_2=a_2 (1-e_2^2)$ is the outer body's semiminor axis, and $e_T$ and $\omega_T$ are those of Triton. If we assume Triton's orbit precesses much more slowly (due to Neptune's $J_2$) than the outer body's (due to Triton and the Sun), we can replace $\omega_T$ with $\Omega_2$, the longitude of the ascending node of the outer body, measured from Triton's pericenter, in Triton's orbital plane (the angles are geometrically identical). The expression above has no dependence on $\omega_2$, so $e=const$. Using Lagrange's eqations, expressions for the evolution of $i_2$ and $\Omega_2$ obey:
\begin{equation}
\label{domdt}
{d \Omega_2 \over d \tau}=-{3 \over 4} \cos i_2 [(1 +{3 \over 2}e_T^2)-{5 \over 2} e_T^2 \cos(2 \Omega_2)],
\end{equation}
\begin{equation}
\label{didt}
{d i_2 \over d \tau}={15 \over 8} e_T^2 \sin i_2 \sin(2 \Omega_2),
\end{equation}
where $\tau=t M_P/(b_2^3 n_T)$. Putting $e_T=0.6$ (which puts Triton close to intersection with Nereid, assuming current pericenters), we numerically integrated Eqs. \ref{domdt} and \ref{didt} for initial conditions of $\Omega_2=90^{\circ}$ and $i_2=0-180^{\circ}$ (Fig. \ref{ioplot}). Suprisingly, objects with $i=40^{\circ}-140^{\circ}$ can have nodes librating around $90^{\circ}$ (or $270^{\circ}$, as the results for $\Omega_2>180^{\circ}$ are symmetric), and their inclination (relative to Triton) can change from direct to retrograde. We tested this model using a numerical integration, in which Triton had its current $i$ and $q$ but $e_T=0.6$. Test particles were placed outside its apocenter, on $e=0$ orbits in Neptune's equator plane. Several particles show librating $\Omega_2$ and repeatedly change their sense of revolution (Fig. \ref{ioplot}). 


Inclinations in Fig. \ref{ioplot} are relative to Triton's orbit, which has $i=157^{\circ}$ relative to Neptune's equator, which itself has an obliquity of $29^{\circ}$ relative to the planet's orbit. Objects in libration island can end up with a wide range of inclinations relative to the Sun. In particular, bodies starting within the equatorial plane of Neptune can be shuffled through the libration island to both direct and retrograde orbits.

One recently discovered irregular, S/2002~N1, has $a=1.6 \times 10^6$~km, $q\geq 1.3 \times 10^6$~km, and $i=134^{\circ}$ \citep{hol04}. \citet{grav04} report that N1's colors are similar to that of Nereid and suggest that they are members of a collisional family. This is implausible if N1 was produced at the present epoch, but is consistent with both objects being pieces of circumplanetary debris ejected by Triton and mixed by this process. 


\section{Summary}

We propose that, after its initial capture by Neptune (via any process), Triton strongly pertubed all of the pre-existing satellites, both regular and irregular. Regular satellites rapidly collide with each other, creating a debris disk to be subsequently swept up swept up by Triton. The debris drag evolved Triton's orbit rapidly enough ($<10^5$~yrs) to preserve some of the irregular satellites, scattering some to high $q$. Bodies in the inner part of the Hill sphere later suffered Triton's secular perturbations, which in some cases ``flipped'' their inclination between direct and retrograde. The closest irregular satellites were subsequently depleted through encounters with Nereid, producing the distribution observed today.

\end{document}